\documentclass[preprint,10pt,5p]{elsarticle}
\pdfoutput=1

\usepackage{atlasphysics} 
\usepackage{xspace}
\usepackage{hyperref}

\newcommand{\rjets}{\ensuremath{R_{\mathrm{jet}}}\xspace}

\newcommand{\ppb}{\ensuremath{\mathrm{pb}^{-1}}\xspace}

\newcommand{\pythia}{{\sc Pythia}\xspace}
\newcommand{\alpgen}{{\sc Alpgen}\xspace}
\newcommand{\herwig}{{\sc Herwig}\xspace}
\newcommand{\jimmy}{{\sc Jimmy}\xspace}
\newcommand{\powheg}{{\sc Powheg}\xspace}
\newcommand{\photos}{{\sc Photos}\xspace}
\newcommand{\tauola}{{\sc Tauola}\xspace}

\DeclareGraphicsExtensions{.pdf}

\journal{Physics Letters B}

\begin{document}

\begin{frontmatter}

\title{\vspace{-0.5cm} \flushright{\normalsize{CERN-PH-EP-2011-126}} \\ \center{ A measurement of the ratio of the \Wboson and \Zboson cross sections with
exactly one associated jet in $pp$ collisions at \rts=7~TeV with ATLAS}}
\author{The ATLAS Collaboration}

\begin{abstract}
The ratio of production cross sections of the \Wboson and \Zboson bosons
with exactly one associated jet is presented as a function of jet transverse
momentum threshold. The measurement has been designed to maximise
cancellation of experimental and theoretical uncertainties, and is reported
both within a particle-level kinematic range corresponding to the detector
acceptance and as a total cross-section ratio. Results are obtained with the
ATLAS detector at the LHC in $pp$ collisions at a centre-of-mass energy of
$7\TeV$ using an integrated luminosity of 33~\ppb. The results are compared
with perturbative leading-order, leading-log, and next-to-leading-order QCD
predictions, and are found to agree within experimental and theoretical
uncertainties.  The ratio is measured for events with a single jet with \pt
$ > 30 \GeV$ to be $8.73 \pm 0.30\, \mathrm{(stat)} \pm
0.40\,\mathrm{(syst)}$ in the electron channel, and $ 8.49 \pm 0.23\,
\mathrm{(stat)} \pm 0.33 \,\mathrm{(syst)}$ in the muon channel.
\end{abstract}


\end{frontmatter}

\section{Introduction}
\label{sec:intro}
Measurements of vector bosons $V$ where $V=W$ or $Z$ produced in association
with one or more jets
($V$+jet)~\cite{CDFWjets:2007,CDFZjets:2008,D0Zjets:2009,ATLASWjets:2010pg}
provide an important test of the Standard Model (SM) description of the
strong interaction in perturbative quantum chromodynamics (QCD).  This is
particularly important in the kinematic region accessible at the LHC in
order to understand the physics at or above the electroweak
symmetry-breaking scale.  Production of vector bosons is also a significant
source of background for studies of other SM processes, including studies of
top quark properties, searches for the Higgs boson, as well as in many
searches for physics beyond the Standard Model.  Measurements of the
kinematic properties and dynamics of $V$+jet processes and comparisons to
theoretical predictions are therefore of significant interest.  Individual
measurements of kinematic observables in $W$+jet~\cite{ATLASWjets:2010pg}
and $Z$+jet events are limited by systematic uncertainties common to
both. Measurement of the ratio was first proposed in
Ref.~\cite{PhysRevD68033014} to exploit the cancellation of theoretical and
experimental uncertainties, therefore building the foundations for a high
precision test of the Standard Model.  In the present measurement, this
ratio is measured, for states involving exactly one jet, as a function of
the minimum jet transverse momentum.  In addition to testing the predictions
of perturbative QCD at various energy scales, the measurement provides
model-independent sensitivity to new physics coupling to leptons and jets.

This Letter describes a measurement of the ratio of the production cross
sections in the electron and muon decay channels of the \Wboson and \Zboson
gauge bosons in association with exactly one jet with transverse momentum
$\pt > 30\GeV$.  The measurement was performed in the active fiducial volume
of the detector and in a kinematic range where events are well-measured, hence
minimising any model-dependence.
The results were corrected to facilitate direct comparison to
theoretical predictions at the particle level.  Following the detector
acceptance, the fiducial regions were defined by $\pT^{\mathrm{lepton}} > 20
\GeV$, $\pT^\nu>25 \GeV$, $\pT^\mathrm{jet} > 30 \GeV$,
$|\eta^\mathrm{jet}| < 2.8$\footnote{ The nominal $pp$ interaction point at
the centre of the detector is defined as the origin of a right-handed
coordinate system.  The positive $x$-axis is directed from the interaction
point to the centre of the LHC ring. The positive $y$-axis points upwards,
while the beam direction defines the $z$-axis. The azimuthal angle $\phi$ is
measured around the beam axis and the polar angle $\theta$ from the
$z$-axis.  The pseudorapidity is defined as $\eta = -\ln
\tan(\theta/2)$}, and for electrons by $1.52 <
|\eta| < 2.47$ or $|\eta| < 1.37$ and for muons by $|\eta|<2.4$.  Events
with a second jet with $\pt > 30\GeV$ within this fiducial region were
rejected.\footnote{This veto is not expected to significantly affect
fixed-order predictions for $V +$ jet cross-sections within the presented
range of jet \pt measurements.}  The selected jet was required to be isolated from
electrons by requiring $\Delta R^{e,\mathrm{jet}} > 0.6$ (where $\Delta R =
\sqrt{(\Delta\eta)^2 + (\Delta\phi)^2}$).  Requirements are made on the
boson masses specific to their reconstruction.  For the \Wboson, the
transverse mass defined by the lepton $\ell$ and neutrino $\nu$ transverse
momenta and angles as $\mT
=\sqrt{2\ensuremath{p_{\mathrm{T}}}^{\ell}\ensuremath{p_{\mathrm{T}}}^{\nu}(1-\cos(\phi^{\ell}-\phi^{\nu}))}$
was required to satisfy $m_{\rm T} > 40 \GeV$.  The dilepton invariant mass
of the \Zboson was required to be within the range $71 < m_{\ell\ell} <
111\GeV$.

Particle-level jets were defined as jets reconstructed in simulated events
by applying the anti-$k_{\rm{t}}$ jet reconstruction
algorithm\cite{Cacciari:2008gp} with a radius parameter $R = 0.4$ to all
final state particles with a lifetime longer than 10~ps (including muons and
non-interacting particles).  Particle-level electrons were defined by
including the energy of all radiated photons within a cone of $\Delta R =
0.1$ around each electron.  The results are compared to
perturbative leading-order~\cite{Pythia}(LO), leading-log~\cite{Alpgen}(LL),
and next-to-leading-order~\cite{MCFM} (NLO) QCD calculations.

\section{The ATLAS Detector}
\label{sec:detector}
The ATLAS detector~\cite{AtlasDetector,CSCbook} consists of an inner
detector tracking system (ID) surrounded by a superconducting solenoid
providing a $2$T magnetic field, electromagnetic and hadronic calorimeters,
and a muon spectrometer (MS).  The ID consists of pixel and silicon
microstrip detectors inside a transition radiation tracker (TRT).  The
electromagnetic calorimeter is a lead liquid-argon (LAr) detector in the
barrel ($|\eta| < 1.475$) and the end-cap ($1.375 < |\eta| < 3.2$) regions.
Hadron calorimetry is based on two different detector technologies. The
barrel ($|\eta| < 0.8$) and extended barrel ($0.8 < |\eta| < 1.7$)
calorimeters are composed of scintillator/steel, while the hadronic end-cap
calorimeters ($1.5 < |\eta| < 3.2$) are LAr/Cu.  The forward calorimeters
($3.1 < |\eta| < 4.9$) are instrumented with LAr/Cu and LAr/W, providing
electromagnetic and hadronic energy measurements, respectively.  The MS is
based on three large superconducting toroids and a system of three stations
of trigger chambers and precision tracking chambers.

\section{Simulated Event Samples}
\label{sec:simulation}
Simulated event samples were used to correct signal yields for
detector effects, for some of the background estimates, and for comparison
of the results to theoretical expectations. Samples of $W \rightarrow
\ell\nu+N_\mathrm{parton}$ and $Z \rightarrow \ell\ell+N_\mathrm{parton}$ (where
$\ell=e,\mu,\tau$) were generated using \alpgen~v2.13~\cite{Alpgen} with the MLM
matching scheme~\cite{MLM}, interfaced to \herwig~v6.510\cite{Herwig} for
parton shower and fragmentation processes, and to
\jimmy~v4.31\cite{Jimmy} for the underlying event simulation.
The CTEQ6L1~\cite{CTEQ6L1} parton density functions (PDFs) were used for
samples generated with \alpgen.
Additional inclusive samples were generated using
\pythia~6.4.21~\cite{Pythia} using the MRST 2007 $\rm{LO}^*$~\cite{LO*} PDF
for the same processes.  The underlying event was generated with the ATLAS
MC09 tune \cite{MC09tune} for the \alpgen and \pythia samples.  A
\powheg~v1.01p4~\cite{Powheg} NLO matrix element calculation with the
CTEQ6.6M~\cite{CTEQ6.6M} PDF set and CTEQ6L1 \pythia parton showering and
underlying event was used to generate \ttbar samples.  The radiation of
photons from charged leptons was treated in \herwig and \pythia using
\photos~v2.15.4~\cite{Photos}, and \tauola~v1.0.2~\cite{Tauola} was used for
tau decays.  The \powheg sample used the ATLAS MC09 tune with one parameter
adjusted.\footnote{The cutoff for multiple parton interactions, PARP(82),
was adjusted from 2.3 to 2.1\GeV, suitable for the CTEQ6L1 PDF.}  Inclusive
samples of charm and bottom quark production were generated with \pythia
6.4.21.  To reproduce the detector conditions, samples were also generated
with multiple inelastic non-diffractive interactions overlaid on top of the
hard-scattering event; the number of additional interactions followed a
Poisson distribution with a mean of approximately two \cite{PV}.  These MC
samples were then re-weighted such that the distribution of the number of
primary vertices matched that of the data.  All samples were passed through
the ATLAS detector simulation~\cite{AtlasSimulation} performed using
GEANT4~\cite{Geant4} and were subjected to the same reconstruction and
analysis chain as the data.

Predictions for the \Wboson + one jet and \Zboson + one jet cross-sections
at NLO were obtained with MCFM~\cite{MCFM} with the same jet algorithm and
kinematic selection requirements as applied to the data.  A correction to
particle level was applied to the MCFM predictions using \pythia to account
for initial and final state radiation, underlying event, and hadronization.
Renormalisation and factorisation scales were set to \HT/2, where \HT~is the
scalar sum of the \pT~of the unclustered partons, the lepton and the
neutrino. The CTEQ6.6M~\cite{CTEQ6.6M} PDF was used for the NLO
calculations.

\section{Data and Event Selection}
\label{sec:selection}
The data used in this analysis were collected in the period March to October
2010.  Basic requirements on beam, detector, stable trigger conditions and
data quality resulted in a data set corresponding to an integrated
luminosity $\mathcal{L} =$ 33~\ppb.  The
criteria for event selection and lepton identification followed 
those employed for the \Wboson and \Zboson inclusive cross-section
measurement~\cite{Aad:2010yt} with a few differences to account for the jet
selection and to optimise cancellation of systematic uncertainties in the
ratio.

In the electron channel, events were selected using a trigger logic that
required the presence of at least one electromagnetic cluster in the
calorimeter with transverse energy $\ET = E \sin(\theta)$ above $15\GeV$ in
the region $|\eta| < 2.5$.  Electron candidates were required to be matched
to a track with silicon pixel and strip measurements in the ID, to have $\ET
> 20 \GeV$, and be within the fiducial region, avoiding the calorimeter
barrel and end-cap transition regions. Candidates were required to satisfy
standard ``tight'' or ``medium'' criteria~\cite{Aad:2010yt}.  Candidates
satisfying lateral shower containment, shape and width criteria with minimal
leakage into the hadronic calorimeter were classified as ``medium''.
Candidates satisfying additional pixel and impact parameter criteria which
also satisfied further requirements on the ratio of cluster energy to track
momentum and on the ratio of high-threshold hits to the total number of TRT
hits were classified as ``tight''.

In the muon channel, events were selected with a trigger system which
identified muon candidates by the presence of hit patterns in the MS,
consistent with a muon track with $\pT > 10\GeV$ or $\pT > 13\GeV$
(depending on the data period). The measured transverse momentum in the MS
was required to satisfy $\pT > 10\GeV$ to reject backgrounds from decays in
flight.  Muon candidates were required to have independent momentum
measurements in both the ID and MS, which were then combined.  Candidates
which satisfied $\pT > 20\GeV$ and were found to be within the fiducial
region were classified as ``medium''.

Muons were additionally classified as ``tight'' if they satisfied all of the
following additional criteria: the impact parameter with respect to the
nominal beam axis was consistent with prompt muon production, the ID track
satisfied additional hit quality criteria, the independent ID and MS track
\pT measurements were consistent, and the muon was isolated
by requiring the $\sum \pt$ of all tracks within $\Delta R~<0.2$ of the muon
to be less than $1.8\GeV$.

Events were required to have at least one reconstructed primary vertex with
three or more associated tracks consistent with the nominal luminous
region. The vertex with the largest $\sum \pT^2$ of associated tracks was
assumed to be the primary vertex and was required to be within $150$~mm
of the centre of the detector along the beam direction.

The missing transverse energy (\met) was calculated from the energy deposits of
calorimeter cells grouped into three-dimensional
clusters~\cite{topoclusters} following the prescription in
Ref. \cite{Aad:2010yt}.  These clusters were corrected to account for the
different response to hadrons compared to electrons or photons, as well as
dead material and energy losses~\cite{lochadtopo}.  The \met was also
corrected for measured muon momenta and their energy depositions in the
calorimeter.  To calculate $m_{\mathrm{T}}$, the $(x,y)$ components of the
neutrino momentum were inferred from the corresponding \met components.

Events containing \Wboson boson candidates were selected by requiring one
electron or muon satisfying ``tight'' selection criteria with no other
``medium'' leptons in the event, $\met > 25\GeV$ and $\mT > 40\GeV$.  Events
containing \Zboson candidates were selected by requiring at least one lepton
satisfying ``tight'' requirements and an additional same flavour, opposite
charge lepton satisfying at least ``medium'' criteria, for which the pair
was required to satisfy $71 < m_{\ell\ell} < 111\GeV$.  The less stringent
requirements on the second lepton reduced the systematic uncertainty on the
lepton identification within the ratio measurement.  These $W$ and $Z$
selections were defined to ensure mutual exclusivity.
		
The jet reconstruction efficiency in simulated data control
samples \cite{JERconf} was found to be close to $100\%$ for jets with $\pT > 30\GeV$.
Events containing jets arising from detector noise or cosmic rays were
rejected \cite{jetcleaning}.  A \pT- and $\eta$-dependent correction factor,
derived from simulated events, was applied to the \pt of each jet to provide
an average energy scale correction~\cite{JES}.  Jets were required to be
reconstructed within $|\eta| < 2.8$ and with \pT~$> 30\GeV$.  To avoid
double-counting of electrons as jets, the closest jet within $\Delta R <
0.2$ of an electron candidate was not considered.  Selected jets were
required to be isolated from selected ``medium'' electrons by requiring
$\Delta R(e,\mathrm{jet}) > 0.6$, to prevent a distortion of the jet energy
response and of the jet reconstruction efficiency due to the proximity of
the electron's electromagnetic shower.  Jets from multiple
interactions in a bunch crossing were suppressed by requiring jets with
associated tracks have a good jet-vertex fraction ($\jvf >
0.75$)~\cite{ATLASWjets:2010pg}.  This algorithm used track-jet association
within a cone of $\Delta R\mathrm{(track,~jet)} < 0.4$.  The \jvf was
computed for each jet as the scalar sum \pT~of all associated tracks which
also originated from the primary vertex divided by the scalar sum \pT of all
the associated tracks.  Events were required to have exactly one jet
satisfying the above criteria with transverse momentum above the jet \pt
threshold. Events containing a second jet with a good \jvf and $\pt >
30\GeV$ were rejected.

Applying these \Wboson (\Zboson) selection criteria, $12112$ ($948$) events
and $12995$ ($1376$) events were retained in the electron and muon channels
respectively.

\section{Background Estimation}
\label{sec:backgrounds}
Two categories of background events were considered, originating from
either QCD multijet or electroweak processes.  The electroweak background
contributions in both channels were estimated from the simulated event
samples as a fraction $f_\mathrm{ewk}$ of the total multijet-subtracted event
yield, which has the advantage that there is no reliance on the measured
absolute luminosity, and reduces the systematic uncertainty from detector
effects on the acceptance.  The multijet contributions were estimated using template
methods based on simulated electroweak and multijet-enriched data
samples, and were also expressed as a fraction $f_\mathrm{multijet}$ of the total
event yield.  The background contributions were derived in the electron and
muon channels for $W$ and $Z$ selections separately for each jet \pt
threshold value.  They were subtracted from the total event yield
$N_\mathrm{tot}$ from Table~\ref{tbl:backgrounds} using,
\begin{eqnarray}
N_\mathrm{sig} & = &   N_\mathrm{tot} \cdot (1-f_\mathrm{multijet})   (1-f_\mathrm{ewk})
\end{eqnarray}
to obtain the signal event yield $N_\mathrm{sig}$.
The yields and breakdown of background predictions  are shown in
Table~\ref{tbl:backgrounds}.

\begin{table}[!htbp]
\begin{center}
\begin{tabular}{l|c|c}
\hline \hline
Process & $W\rightarrow e \nu$  & $Z\rightarrow e e$  \\
\hline 
 $W\rightarrow e \nu$   & $9340 \pm 40 $ & $3 \pm 1$ \\
 $Z \rightarrow ee$   & $106 \pm 3$ & $880 \pm 10$ \\
 $W\rightarrow \tau \nu$  & $191 \pm 6$ & $ 0.2 \pm 0.2 $ \\
 $t\bar{t}$ & $33\pm 1$ & $1.9 \pm 0.2$ \\
 $Z \rightarrow \tau\tau$  & $ 19 \pm 1 $ & $0.3 \pm 0.1$\\
Multijet  & $1800 \pm 60$ & $2.9 \pm 0.6$ \\
Other & $58 \pm 2$  & $1.1 \pm 0.1$ \\
\hline
 Total & $11550 \pm 70$ & $ 880 \pm 10$ \\
\hline
Data $N_\mathrm{tot}$      &  $12112$  &    $948$   \\
\hline \hline
Process & $W\rightarrow \mu \nu$  & $Z\rightarrow \mu\mu$  \\
\hline
$W \rightarrow \mu \nu$ & $11860 \pm 40$ & $ 4 \pm 2$ \\
$Z \rightarrow \mu \mu$ & $360 \pm 6 $ & $1370 \pm 40$ \\
$W \rightarrow \tau \nu$ & $234 \pm 6$ & $0.3 \pm 0.6$ \\
$Z \rightarrow \tau \tau$ & $22 \pm 1$ & $0.3 \pm 0.6$\\
$t \bar{t}$ & $35 \pm 1$ & $3 \pm 2$ \\
Multijet  & $380 \pm 70$ & $4 \pm 4$ \\
Other & $117 \pm 1$ & $8 \pm 3$ \\
\hline
Total & $13010 \pm 80$ & $1380 \pm 40$ \\
\hline 
Data  $N_\mathrm{tot}$ & $12995 $ & $1376$ \\
\hline \hline
\end{tabular}
\caption{Predicted and observed event yields in data in the
electron and muon channels for the $W$ and $Z$ selections for $\mathcal{L} =
33$~\ppb.  Background estimates are quoted for a jet \pt threshold of
$30\GeV$.  ``Other'' includes contributions from diboson and single top
events.  The total statistical uncertainties on predictions are quoted.}
\label{tbl:backgrounds}
\end{center}
\end{table} 

The background contribution from multijet processes in the electron channel
originates from events with jets misidentified as electrons and
the mismeasurement of calorimetric energy resulting in large \met. This
background contribution was estimated by using a partially data-driven
method~\cite{Aad:2010yt}.

The multijet background within the $W \rightarrow e\nu$ sample selection was
estimated by fitting templates to the low \met control region $ 15 < \met <
55\GeV$.  The \met templates for signal and electroweak processes were
derived from Monte Carlo simulations, while the template for the multijet
contribution was extracted from data by inverting the ``tight'' electron
identification criteria which are not correlated with the \met.  The result
of this fit was the relative contribution of the multijet
background~\cite{Aad:2010yt} to the data. This estimate was performed for
each jet \pt threshold considered.
 
The multijet background in the \Zee\ channel was estimated with a similar
fit using the di-electron invariant mass distribution.  Templates for signal
and electroweak processes were derived from Monte Carlo simulated events, while
the template describing the multijet contribution was obtained by inverting
two of the ``medium'' selection criteria of the \Zboson selection.

For the $W$ selection in the electron channel, the electroweak contributions
mainly originate from $W \rightarrow \tau\nu$ events where the $\tau$ decays
to an electron, and \ttbar where one or more \Wboson decay to an electron.
For the $Z$ selection, they similarly come from \ttbar events, and from
\Ztau where both $\tau$ leptons decay to electrons.  Finally,
\Wboson and \Zboson production also constitute significant background to
each other, due to events in which one electron from the $Z$ was not
reconstructed, or when a \Wboson event contains an additional electron
candidate.  The total electroweak background fraction in the electron
channel was approximately $3.4\%$ for the selected $W$
candidates, and less than $1\%$ for the $Z$ candidates.

The multijet background to $\Wboson \rightarrow \mu \nu +$ jet events 
is estimated from the number of events passing all signal selections except
isolation, and efficiencies derived from control samples in data for signal
and multijet events required to pass the isolation requirement.

To estimate the multijet backgrounds for the \Zboson in the muon channel,
non-isolated muon pairs were selected in the simulated multijet sample.  By
comparing the dimuon invariant mass distribution for this sample and a
non-isolated sample in data, a scale factor was derived that was used to
normalise a simulated background sample with the isolation requirement
applied.

For the \Wboson selection in the muon channel, the electroweak backgrounds
mainly originate from decays \Wtau with the $\tau$ decaying to a muon and
\Zmm where one muon fails to be reconstructed.  The \Ztau and \ttbar processes
where one or both \Wboson boson(s) decay to a muon, contribute a smaller
background fraction.  For the \Zboson selection in the muon channel, the
dominant electroweak backgrounds arise from \Ztau and \ttbar events with two
real muons in the final state.  The total electroweak background fraction
in the muon channel was approximately $5\%$ for the selected $W$
candidates, and less than $1\%$ for the $Z$ candidates.

\section{Correction to Particle-level Yield}
\label{sec:signal}
The number of events after selection and background subtraction
($N_\mathrm{sig}^{\ell,V}$) for each lepton $\ell$ and boson $V$ was
corrected for the detector effects and selection efficiencies back to the
particle level ($N_\mathrm{part}^{\ell,V}$).  This corrected yield can be
directly compared to theoretical predictions at the particle level.
The corresponding correction factors from detector to 
particle level were computed for each jet \pt threshold  
in the electron and muon channels for the $W$ and $Z$ selections separately.

Yields were corrected with multiplicative factors which include trigger
efficiency ($\epsilon_{trig}^{\ell}$), lepton identification efficiency
($\epsilon^\ell$), and boson reconstruction and resolution ($C_{V}^{\ell}$).
The number of signal events for each boson at particle level was then
obtained using
\begin{eqnarray}
\label{eqn:e-had}
{N_\mathrm{part}^{\ell,V}} &=& \frac{N_\mathrm{sig}^{\ell,V}} { \epsilon_\mathrm{trig}^\ell \times \epsilon^\ell
\times C_{V}^{\ell}}.
\end{eqnarray}
where the boson corrections $C_{V}^\ell$ correct the observed phase space to
the fiducial phase space defined above, accounting for the resolution of
leptons and \met.

Trigger and identification efficiencies were binned to account for
variations in detector response.  These efficiencies were binned in $E_T$
and $\eta$ for electrons and $\eta$ and $\phi$ for muons.  Efficiencies were
found to be independent of the jet multiplicity and jet \pt. Therefore, a
single efficiency map was used for all jet \pt thresholds.

The methods used to derive these efficiencies and corrections were similar for
the electron and muon channels.  The trigger efficiency and identification
efficiency were measured using a sample of unbiased leptons obtained by
selecting a well identified tag lepton in $Z \rightarrow \ell\ell$ candidate
events.  The boson reconstruction correction $C_{V}^{\ell}$ was computed
using the \alpgen event generator.  The \pythia Monte Carlo, used for
comparison, was found to produce a consistent correction factor.

\begin{figure}
\centering
\includegraphics[width=0.45\textwidth]{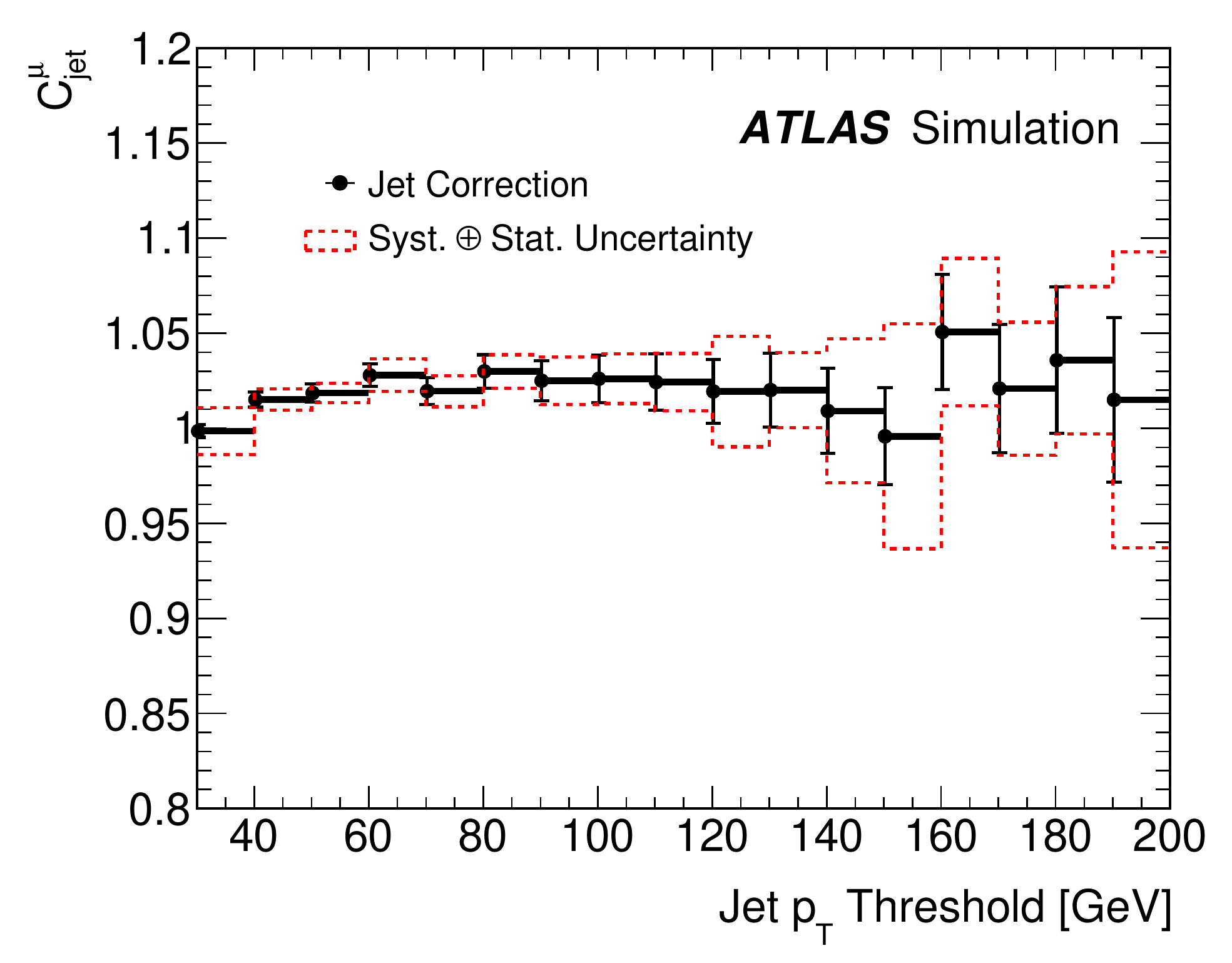}
\caption{
Detector jet spectrum correction ($C_\mathrm{jet}^{\mu}$) on \rjets in the
muon channel derived from \alpgen. The uncertainty is shown as a dashed red
line, it accounts for the difference between \pythia and \alpgen
generators.}
\label{fig:jet-correction}
\end{figure}

By measuring the ratio, almost complete cancellation of jet resolution
effects was achieved.  A small correction $C_\mathrm{jet}^\ell$
was applied to the ratio to account for remaining non-cancelling effects due
to lepton selection, jet selection criteria, and isolation criteria.
The ratio measured was then the ratio of yields corrected to particle level
and finally corrected for these remaining effects:
\begin{eqnarray}
\rjets &=&
\frac{
N_\mathrm{part}^{\ell,W}
}{
N_\mathrm{part}^{\ell,Z}
}
\times C^\ell_\mathrm{jet}.
\end{eqnarray}
The jet correction for the muon channel $C_\mathrm{jet}^{\mu}$ is shown in
Fig.~\ref{fig:jet-correction}.  Systematic uncertainties from this
correction were evaluated on the ratio itself.  The jet correction
$C_\mathrm{jet}^{\ell}$ accounts for the difference of the ratio when
calculated in terms of jets defined at particle level and reconstructed
jets.  The correction factor is different from unity if an offset exists
between $W$+jet and $Z$+jet events in the jet \pt migration from particle
level to detector level.  This offset is due to the different requirements
applied in the $W$+jet and $Z$+jet selections prior to the jet selections,
placing the jets into slightly different phase space regions for the
numerator and the denominator of the measurement.  Performing the
measurement as a function of \pt threshold instead of differentially removes
the effects of migration across the upper bin edge.

\section{Systematic Uncertainties}
\label{sec:systematics}
\begin{figure*}[!htbp]
  \begin{minipage}{0.5\linewidth}
  \begin{center}
    \includegraphics[width=0.95\linewidth]{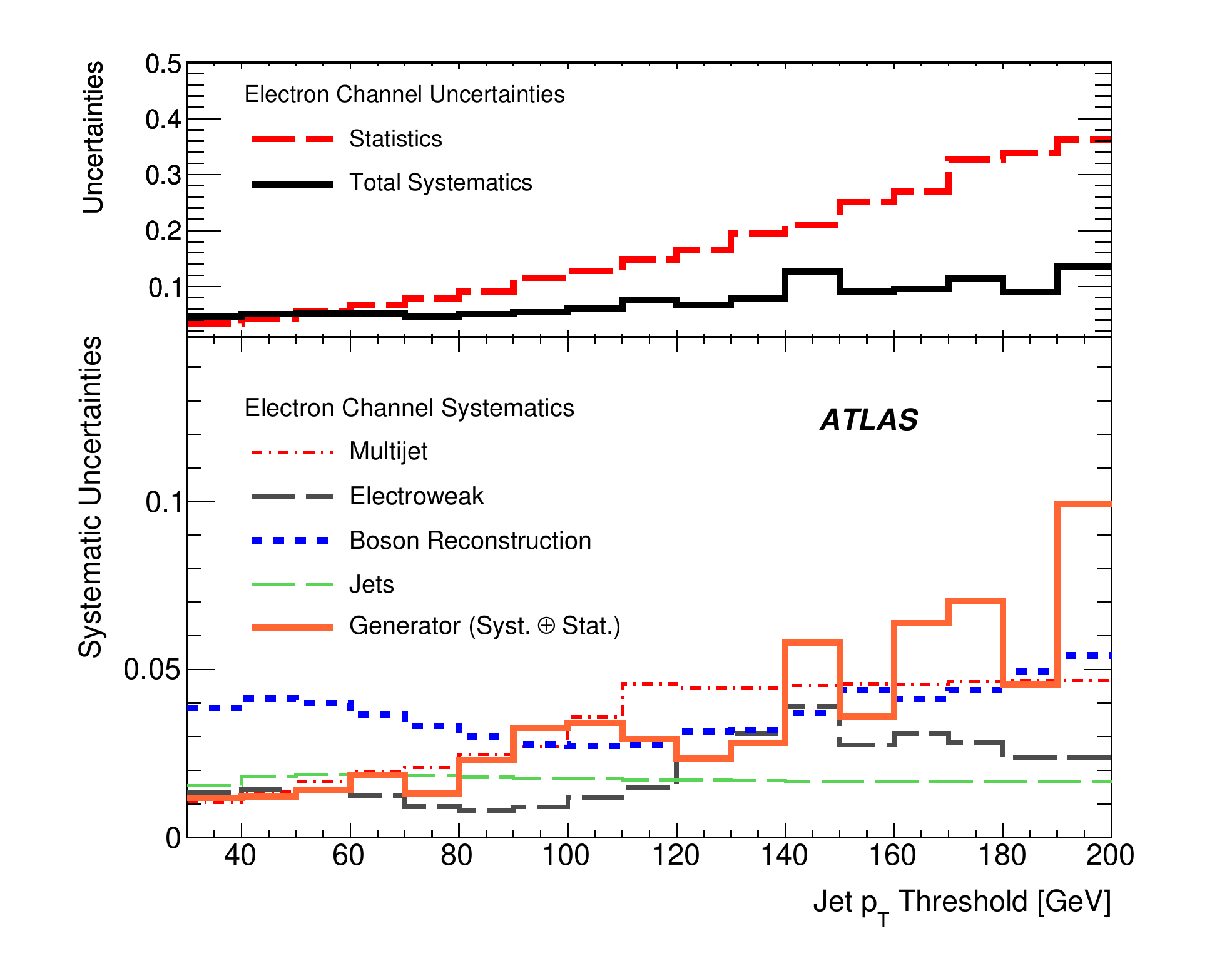}
\end{center}
\end{minipage}
  \begin{minipage}{0.5\linewidth}
    \begin{center}
    \includegraphics[width=0.95\linewidth]{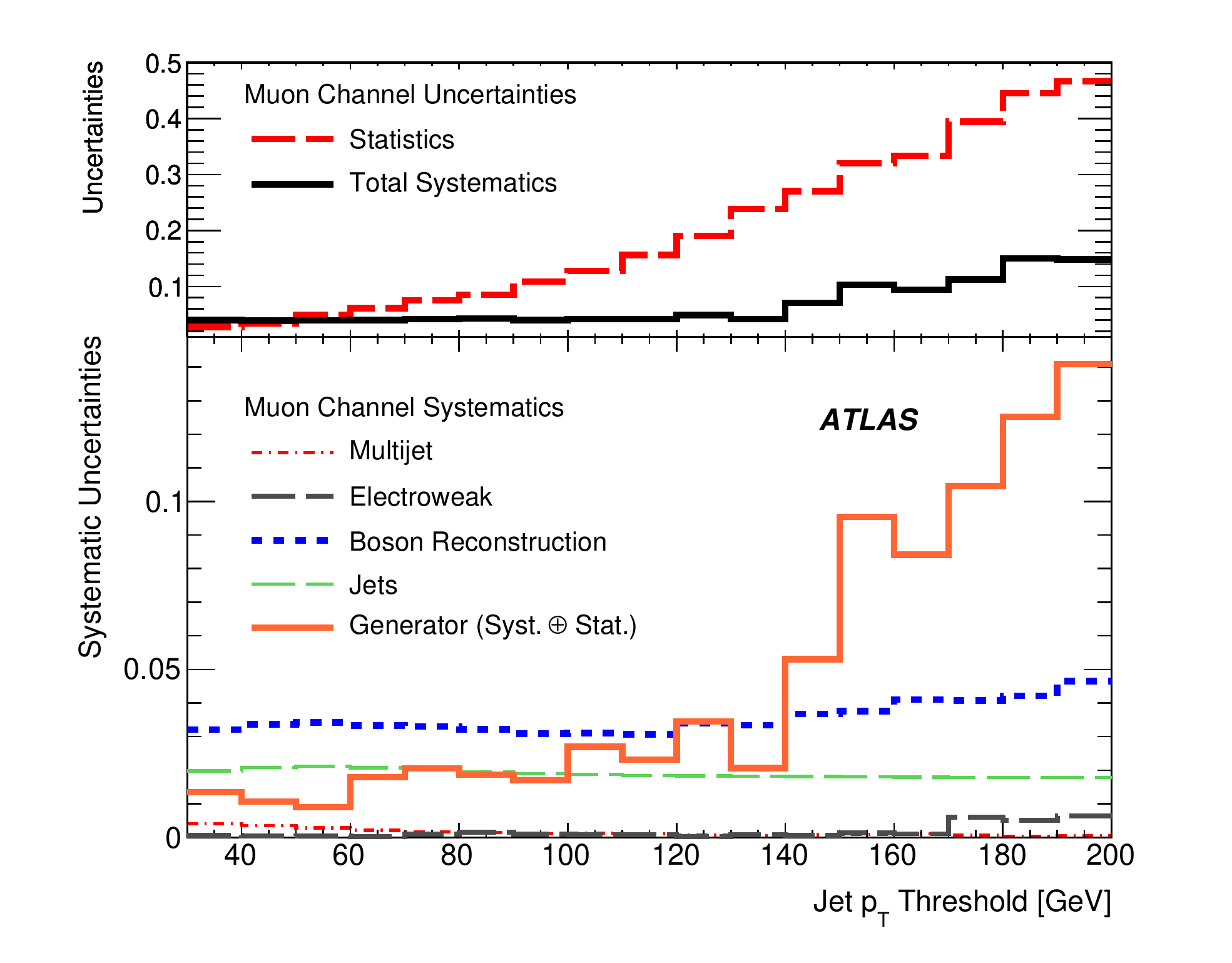}
  \end{center}
  \end{minipage}
\caption{
Relative systematic uncertainties on \rjets in the electron channel (left)
and in the muon channel (right). The top plot displays the total systematic
and statistics uncertainty (shown as red dashed line) versus jet \pt
threshold. The lower plot shows the breakdown of the systematic
uncertainties. Boson reconstruction contains the uncertainties related with
the leptons and \met (including trigger and lepton identification). Jets
contains systematics of the jet correction as well as the jet energy scale
and resolution. Uncertainties from each group were added in quadrature.}
\label{fig:systematics}
\end{figure*}

To evaluate cancellations of systematic uncertainties which occur in the
ratio, the correlations between \Wboson and \Zboson systematic effects must
be considered.  Correlations between the measurements at each jet \pt
threshold must also be accounted for.  The effects of systematic
uncertainties were therefore evaluated by measuring the relative change in
the ratio \rjets from each source.

The total systematic uncertainty ranges from 4\% at low jet \pt to 15\%
for the largest \pt threshold studied. For jet \pt thresholds of greater
than $50\GeV$ the statistical uncertainty dominates the total measurement
uncertainty.

The sources of systematic uncertainties on \rjets were grouped into
uncertainties on the boson reconstruction (including lepton trigger,
reconstruction and identification efficiencies, as well as lepton and \met
scales and resolutions), on jet-related corrections, multijet and electroweak
background predictions, and generator-related uncertainties.

In the muon channel, where the background was small, the
uncertainty on \rjets from the background estimation was approximately 1\%
for the whole jet \pt range. In the electron channel, the uncertainty
increases as a function of jet \pt threshold.  This is due to the larger
background in the electron channel and the limited statistics used to
compute backgrounds for high jet \pt thresholds.

Systematic uncertainties on the multijet background fractions were estimated
by varying the criteria used to derive the background fractions.  Each
systematic uncertainty includes a component from the statistical uncertainty
on the estimate of the background fraction.

The estimate of the electroweak background is affected by systematic effects
from the event selection criteria.  Samples with and without multiple $pp$
interactions included in the simulation were also compared.

The lepton trigger efficiency and identification uncertainties were
estimated following the procedure documented in Ref.~\cite{Aad:2010yt}.  The
uncertainty on the ratio from lepton identification efficiencies was
directly obtained by scaling the single lepton identification efficiencies
by their uncertainties, taking cancellations into account.  A contribution
to the uncertainty on the identification efficiency was assigned from the
difference between its value derived in data and Monte Carlo. The total
identification uncertainty was $1.1\%$ ($1.7\%$) for electrons (muons)
independent of jet \pt thresholds.

The uncertainties on the scale and resolution of lepton energies and \met
were propagated to evaluate their effects on boson reconstruction by
smearing the simulated signal samples using a Gaussian with a width
corresponding to the nominal uncertainties. The resulting variations in
\rjets were applied as systematic uncertainties.

Uncertainties on the jet energy scale (JES) and jet energy resolution (JER)
were determined by comparing data and simulations~\cite{JES}.  The JES
uncertainty includes components from calibration and jet sample composition
differences.  The JES calibration uncertainty varies with $|\eta|$ and \pt,
and ranges from 4\% to 8\%.  The JES and JER were measured with di-jet
events, which have different proportions of quark and gluon initiated jets
than events containing vector bosons.  Therefore, an uncertainty was assigned
to account for the difference in calorimeter response between jets in $V$ +
jet events and the di-jet events used for calibration, ranging from 2 to
5\%, and was added in quadrature to the JES calibration uncertainty.  The
total JES uncertainty ranges from approximately 10\% at 20\GeV to 5\% at
100\GeV.

To compute the effect of the JER uncertainty on the ratio \rjets, jets were
smeared according to a Gaussian with a width corresponding to the JER. The
effect of the JES uncertainty on the ratio was obtained in a similar manner,
but in this case, shifting the jet energy by its uncertainty. The ratio was
recomputed applying these variations simultaneously to the numerator and
denominator.  The change was applied as a systematic uncertainty.  The
uncertainties on
\rjets due to the JER and JES were approximately 0.5\% and 2\%
respectively.  The contribution to the uncertainty on the
ratio from the small component of heavy flavour jets is covered by the total
JES uncertainty.

To account for systematics associated with the modeling of the signal at
particle level, correction factors were re-computed with samples generated
with \pythia instead of \alpgen, and the observed variation was applied as a
systematic uncertainty. Systematic uncertainties were assigned from this
variation to the following corrections: ($C_\mathrm{jet}^\ell$), the boson
reconstruction correction $C_{V}^{\ell}$, and the electroweak background
estimation $f_\mathrm{ewk}$.  At large jet \pt threshold, where the
statistical uncertainty on the measurement dominates the total uncertainty,
this systematic uncertainty is limited by the statistics of the samples used
and is the dominant systematic uncertainty.

The uncertainties due to multiple $pp$ interactions are dominated by
uncertainties on the efficiency of the \jvf algorithm.  It was confirmed
that the results obtained with simulated signal samples which include this
effect were consistent with those obtained from samples which contained no
additional interactions in the simulation.  The residual difference on
\rjets between samples with and without multiple interactions included was
used as the systematic error from this \jvf requirement.

Corrections to the simulation for hadronisation and the underlying event on
the NLO parton-level calculation were computed with \pythia as a function of
jet \pT threshold.  The impact of this correction on \rjets was 1 to 6\% for
the electrons and 1 to 4\% for the muons. The slightly larger variation for
the electrons was due to the jet-electron isolation and the jet isolation
veto included in the corrections.

The uncertainty on the correction of the MCFM cross-section ratio
predictions for fragmentation, hadronisation and underlying event effects
was estimated by comparing the \pythia AMBT1~\cite{AMBT1} tune with the
AMBT1 tune with increased underlying event activity, and without any
underlying event.  The uncertainty due to initial and final state radiation
(ISR/FSR) was evaluated by varying the \pythia parameters controlling ISR
and FSR~\cite{Pythia}. For the ISR the variation ranges used were similar to
the ranges used in the Perugia Soft and Perugia Hard tunes. For the FSR the
variation ranges were similar to the ranges used in the Perugia 2011 radHi
and Perugia 2011 radLo tunes~\cite{Skands:2010ak}.

Renormalisation and factorisation scale uncertainties were estimated by
varying the scales in all combinations, up and down, by a factor of two.
Although these variations are arbitrary, they are motivated by the
dependence of the behaviour of
the NLO W+jet and Z+jet cross section on the scale.  This
choice has a minimal impact on the uncertainty of the \rjets prediction.
Systematic uncertainties from imperfect knowledge of PDFs were computed by
summing in quadrature the dependence on each of the 22 eigenvectors
characterising the CTEQ6.6 PDF set; the uncertainty in $\alpha_{\rm{s}}$ was
also taken into account. An alternative PDF set, MSTW2008 \cite{MSTW2008},
with its set of 68\% C.L. eigenvectors was also examined, and the envelope of
the uncertainties from CTEQ6.6 and MSTW2008 was used as the PDF uncertainty.

All systematic uncertainties were added in quadrature to obtain the total
systematic uncertainty.  A summary of sources of systematic uncertainties and
their relative contributions to \rjets is shown in
Figure~\ref{fig:systematics} and in Table~\ref{tab:systematics}.

\begin{table} 
\begin{center} 
\begin{tabular}{lrcc} 
\hline \hline
\multicolumn{4}{l}{Systematic Uncertainties on \rjets (\%)}  \\
\hline
Electron Channel & $\pt > $ & $30\GeV$ & $100\GeV$  \\
\hline
\multicolumn{2}{l}{Trigger} &  $0.5$ & $0.5$  \\
\multicolumn{2}{l}{Boson Reconstruction} &  $3.7 $ & $2.4$  \\
\multicolumn{2}{l}{Identification Efficiency} &  $1.1$ & $1.1$  \\
\multicolumn{2}{l}{\jvf}      &  $0.0$ & $0.6$  \\
\multicolumn{2}{l}{JES/JER}                   &  $1.5$ & $1.7$  \\
\multicolumn{2}{l}{Multijet Background} &  $1.0$ & $3.6$  \\
\multicolumn{2}{l}{Electroweak Background} &  $1.3$ & $1.2$  \\
\multicolumn{2}{l}{Total Non-Generator} &  $4.5$ & $5.0$  \\
\hline
\multicolumn{2}{l}{Generator } &  $1.2$ & $3.4$  \\
\hline
\multicolumn{2}{l}{Total} &  $4.6$ & $6.1$  \\
\hline\hline 
Muon Channel & $\pt >$ & $30\GeV$ & $100\GeV$  \\
\hline
\multicolumn{2}{l}{Trigger} &  $1.0$ & $1.0$  \\
\multicolumn{2}{l}{Boson Reconstruction}  &  $2.5$ & $2.4$  \\
\multicolumn{2}{l}{\jvf} &  $1.7$ & $0.2$  \\
\multicolumn{2}{l}{JES/JER} &  $2.0$ & $1.9$  \\
\multicolumn{2}{l}{Multijet Background} &  $0.4$ & $0.1$  \\
\multicolumn{2}{l}{Electroweak Background} &  $0.1$ & $0.1$  \\
\multicolumn{2}{l}{Total Non-Generator} &  $3.8$ & $3.2$  \\
\hline
\multicolumn{2}{l}{Generator} &  $1.4$ & $2.7$  \\
\hline
\multicolumn{2}{l}{Total} &  $4.0$ & $4.2$  \\
  \hline \hline
  \end{tabular}
\caption{
Summary of the systematic uncertainties on \rjets for the electron and muon
channels. The value of each of the systematics for jet \pt thresholds of 30
and $100\GeV$ is shown.  For the muon channel, the identification efficiency is
included in the Boson Reconstruction category.}
\label{tab:systematics}
\end{center}
\end{table}

\section{Results}
\label{sec:results}
The ratio \rjets was measured in the fiducial region of the ATLAS detector
defined by the jet \pt threshold and selection criteria.  The present
measurements were corrected back to the particle level and within the
defined kinematic range. The electron and muon measurements were performed
in slightly different phase space, due to the different $\eta$ range and
electron-jet isolation requirements, as well as for the different QED
treatment between electron and muon definitions.  The observed signal yields
were corrected to recover the yield at particle level as described in
Section~\ref{sec:signal}.

\begin{figure*}[!htbp]
  \begin{minipage}{0.5\linewidth}
  \begin{center}
    \includegraphics[width=0.95\linewidth]{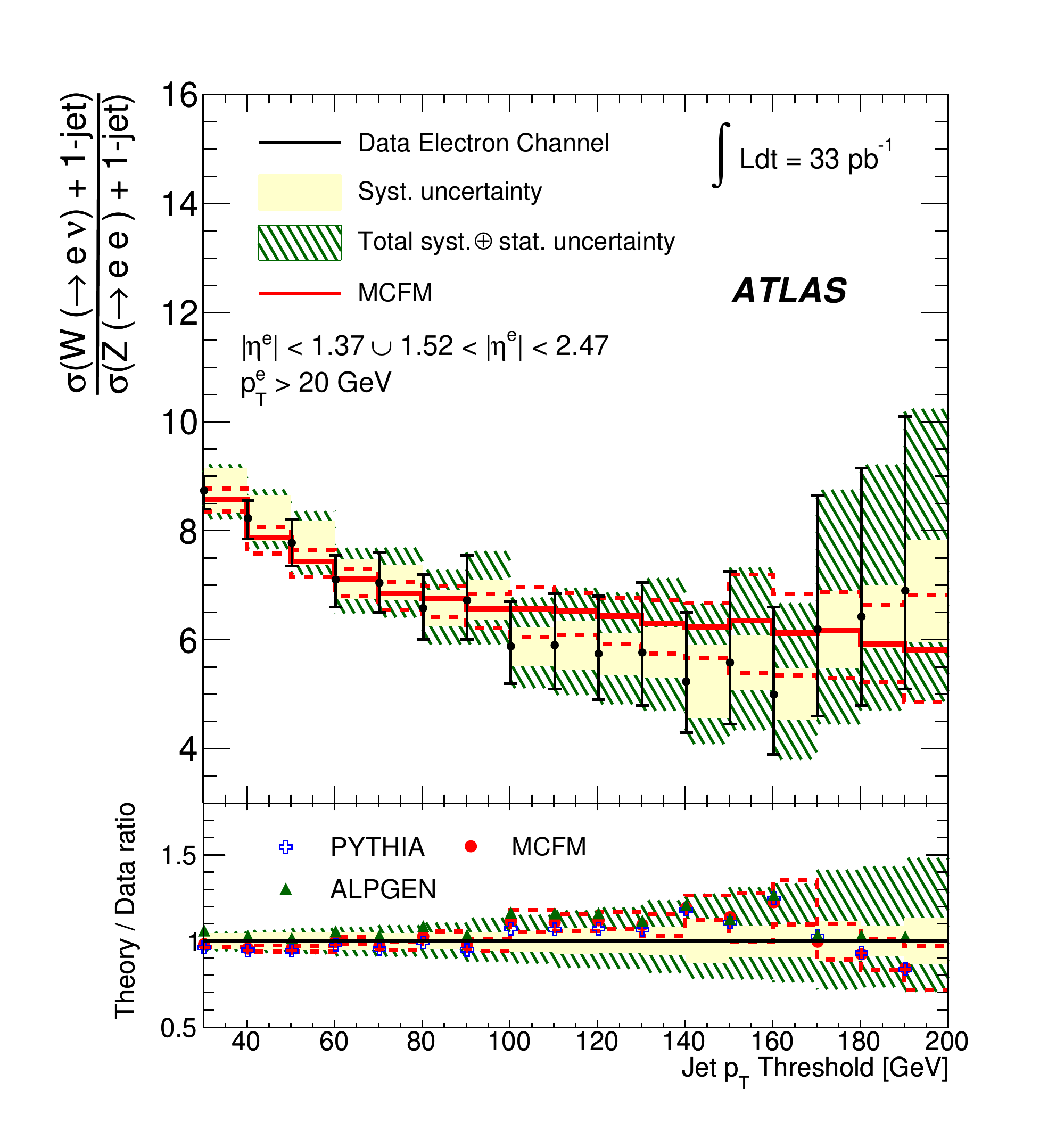}
  \end{center}
  \end{minipage}
  \begin{minipage}{0.5\linewidth}
    \begin{center}
    \includegraphics[width=0.95\linewidth]{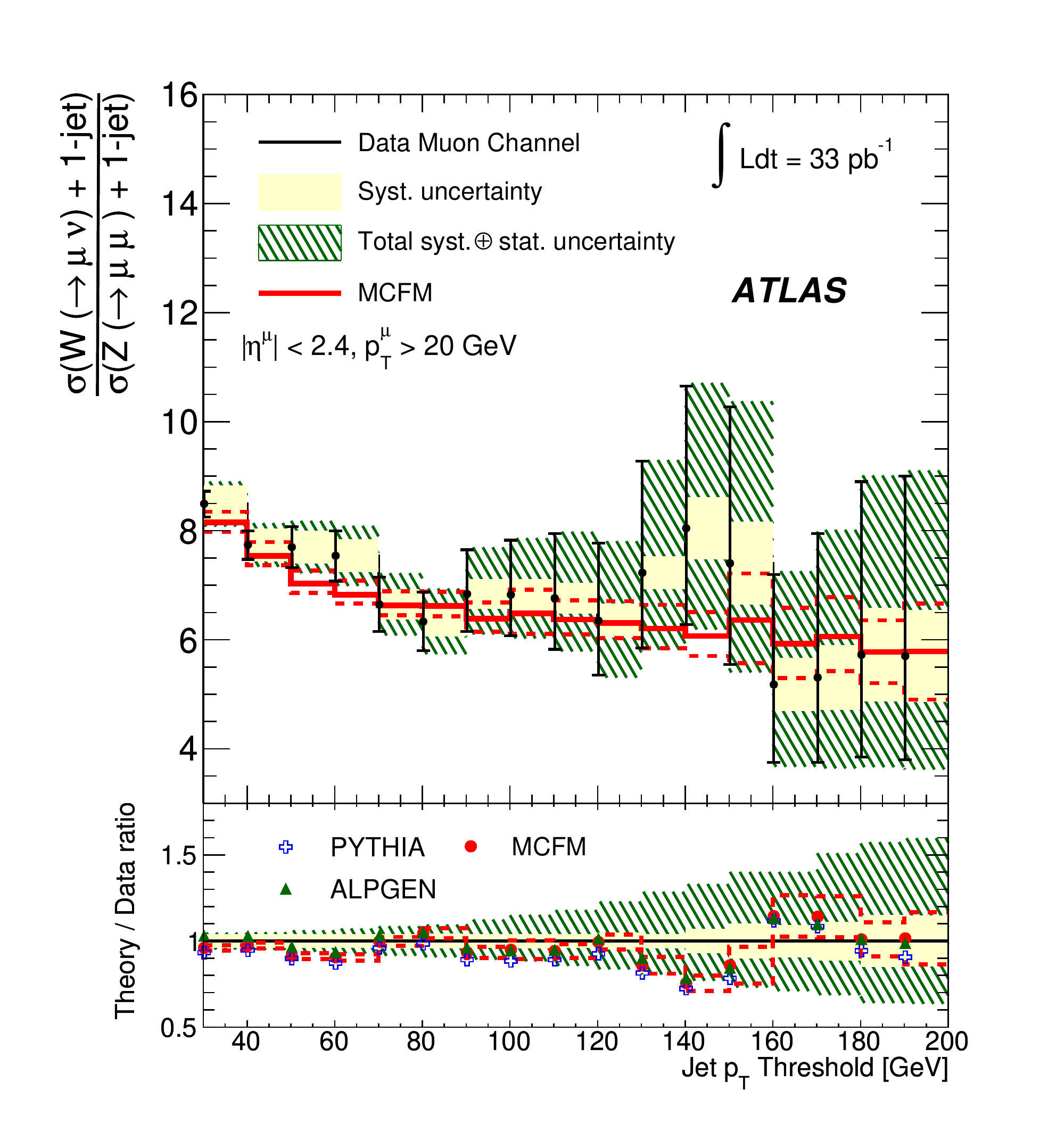}
  \end{center}
  \end{minipage}
\caption{
Results for \rjets in the electron channel (left) and in the muon channel
(right) for their respective fiducial regions.  The results are compared to
NLO predictions from MCFM (corrected to particle level using \pythia). Data
are shown as black points at the lower bin edge corresponding to the jet \pt
threshold with black error bars indicating the statistical uncertainties.
The yellow band shows all systematic uncertainties added in quadrature and
the green band shows statistical and systematic uncertainties added in
quadrature. The theory uncertainty (dashed line) shown on the MCFM
prediction includes uncertainties from PDF and renormalisation and
factorisation scales.  Note that these threshold data and their associated
uncertainties are correlated between bins.}
\label{fig:results}
\end{figure*}

The corrected ratio \rjets of the production cross sections in the leptonic
(electron or muon) decays of the gauge bosons $W$ and $Z$ in association
with exactly one jet is shown in Figure~\ref{fig:results} as a function of
the jet \pT threshold for the electron (left) and muon (right) channels.
As the jet \pt threshold increases, the ratio \rjets is expected to decrease
as the effective scale of the interaction becomes large compared to the
difference in boson masses.  This dependence is observed in the data.
The values for the lowest jet \pt threshold of $30\GeV$ are:

\vspace{-2em}
\begin{center}
\begin{eqnarray*}
\rjets(e)&=& 8.73 \pm 0.30 \,\mathrm{(stat)} \pm 0.40 \,\mathrm{(syst)}  \\
\rjets(\mu)&=& 8.49 \pm 0.23 \,\mathrm{(stat)} \pm 0.33 \,\mathrm{(syst)}
\end{eqnarray*}	
\end{center}
The statistical uncertainties were evaluated by repeating the measurement with
Monte Carlo pseudo-experiments assuming Poisson distributed data with a mean
at the observed yield.  Both electron and muon channel
results are individually compatible with the theoretical predictions.

\begin{figure*}[!htbp]
  \begin{minipage}{0.5\linewidth}
  \begin{center}
  \includegraphics[width=0.95\linewidth]{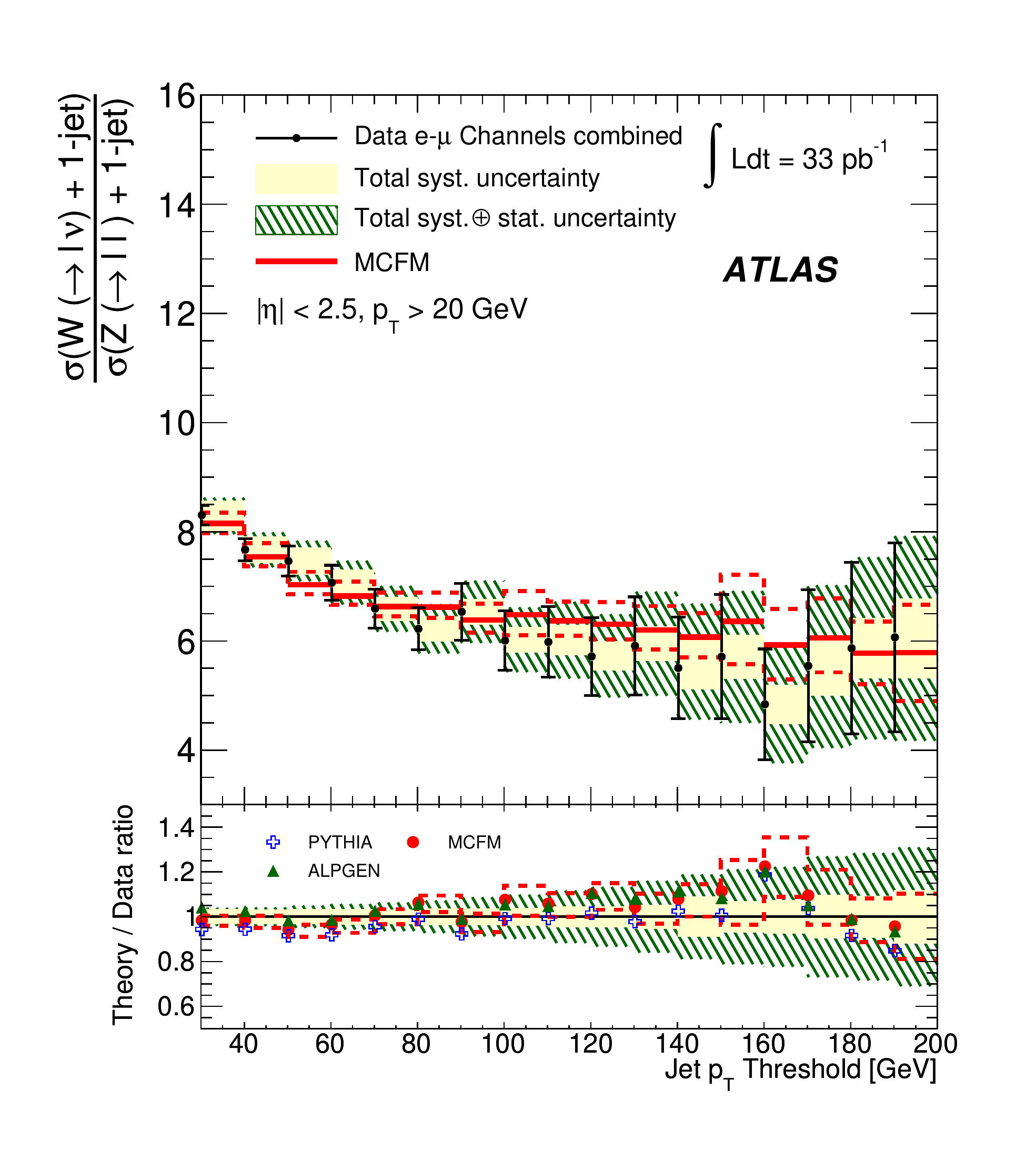}
  \end{center}
  \end{minipage}
  \begin{minipage}{0.5\linewidth}
  \begin{center}
    \includegraphics[width=0.95\linewidth]{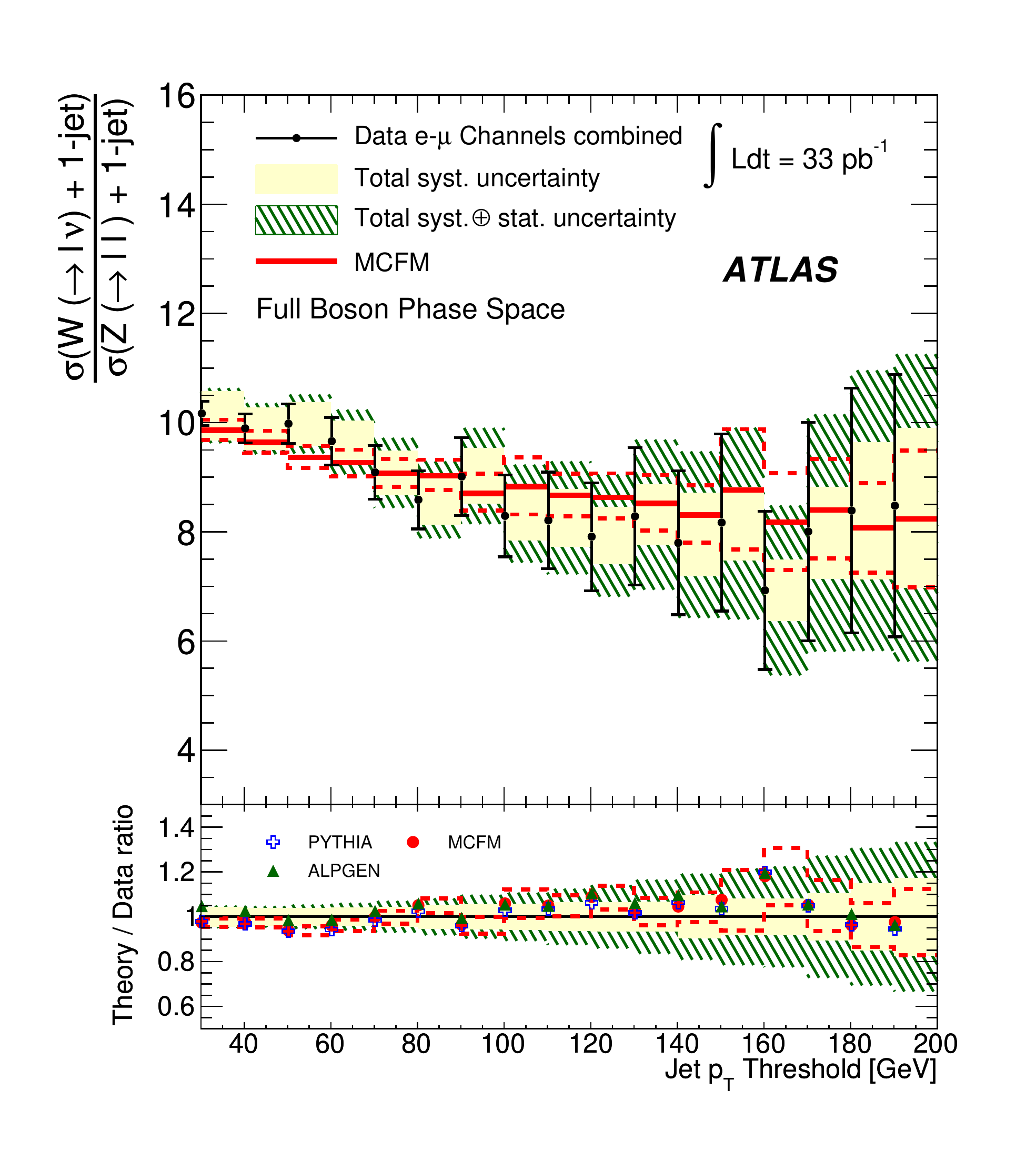}
  \end{center}
  \end{minipage}
\caption{
Left: Combined electron and muon results for \rjets in a common fiducial
region.  The results are compared to predictions from MCFM (corrected to
particle level).  Data are shown with black error bars indicating the
statistical uncertainties. The yellow band shows all systematic
uncertainties added in quadrature and the green band shows statistical and
systematic uncertainties added in quadrature.  The theory uncertainty
(dashed line) includes contributions from PDF and renormalisation and
factorisation scales.  Right: Combined electron and muon results for \rjets
extrapolated to the total phase space.  Note that these threshold data and
their associated uncertainties are correlated between bins.}
\label{fig:combination}
\end{figure*}

Electron and muon channel results were compatible and were therefore
combined to reduce the statistical and uncorrelated systematic uncertainties
on the result. Each channel was extrapolated to a common phase space,
defined as $|\eta_\ell| < 2.5$ before any QED radiation (Born level) with
\pythia.  The electron channel was further
corrected for the effect on the acceptance of the electron-jet isolation
requirements.  This extrapolation to a common fiducial region decreases the
value of the ratio for both channels primarily due to the more central
distribution of leptons from the \Zboson.  The results were combined using a
Bayesian approach~\cite{Caldwell:2010zz} in the combination of systematic
uncertainties accounting for correlations between them.  The systematic
uncertainties from \met, jet energy scale and resolution and electroweak
background sources were considered fully correlated between the electron and
muon channels.  The combined result is shown in
Figure~\ref{fig:combination}~(left).  The value of \rjets for the lowest jet
\pt threshold of 30\GeV is found to be $8.29 \pm 0.18 \,\mathrm{(stat)} \pm 0.28 \,\mathrm{(syst)}$.

The combined results were also extrapolated to the full phase space,
correcting for regions of geometric and kinematic acceptance not measured
(only the requirement on the invariant mass of the $Z$ and veto on
additional jets were retained), and presented as a function of jet \pt
threshold.  For this extrapolation, the \Wboson and \Zboson acceptance
factors were calculated at particle level using
\alpgen, and their ratio was applied as a correction to \rjets.
This extrapolation to the full phase space increases the value of the ratio
for both channels primarily due to additional kinematic acceptance for the
\Wboson.  Figure~\ref{fig:combination}~(right) shows the result in the full
boson acceptance phase space for the combined electron and muon channels.
Due to the additional uncertainty on the correction to the generator-level
acceptance, this result has larger total uncertainty than the results
obtained in the fiducial regions.  No significant discrepancy between data
and theory is observed in any of these results.

\begin{table*} 
\begin{center} 
\begin{tabular}{| c | rcl |  rcl | rcl | rcl |} 
\hline \hline
jet \pt & \multicolumn{3}{c|}{Electron Fiducial} & \multicolumn{3}{c|}{Muon
Fiducial}   &  \multicolumn{3}{c|}{Combined}        &
\multicolumn{3}{c|}{Boson Full Phase Space}  \\
Threshold & \multicolumn{3}{c|}{$|\eta^e|<2.47$}    & \multicolumn{3}{c|}{$|\eta^\mu|<2.4$} & \multicolumn{3}{c|}{$|\eta^\ell|<2.5$} & \multicolumn{3}{c|}{}\\
(\GeV) & \multicolumn{3}{c|}{(excl. $1.37<\eta<1.52$)} & \multicolumn{3}{c|}{}& \multicolumn{3}{c|}{}&\multicolumn{3}{c|}{} \\
\hline
30  & 8.73 &$\pm$ 0.3 &$\pm$ 0.40    & 8.49  &$\pm$ 0.23  &$\pm$ 0.33   & 8.29  &$\pm$ 0.18  &$\pm$ 0.28  & 10.13  &$\pm$ 0.22  &$\pm$ 0.45 \\
40  & 8.23 &$\pm$ 0.35 &$\pm$ 0.41   & 7.74  &$\pm$ 0.26  &$\pm$ 0.30   & 7.67  &$\pm$ 0.20  &$\pm$ 0.24  &  9.89  &$\pm$ 0.26  &$\pm$ 0.38 \\
50  & 7.77 &$\pm$ 0.42 &$\pm$ 0.39   & 7.7   &$\pm$ 0.37  &$\pm$ 0.30   & 7.46  &$\pm$ 0.27  &$\pm$ 0.25  &  9.97  &$\pm$ 0.36  &$\pm$ 0.39 \\
60  & 7.10 &$\pm$ 0.47 &$\pm$ 0.36   & 7.54  &$\pm$ 0.46  &$\pm$ 0.30   & 7.07  &$\pm$ 0.32  &$\pm$ 0.24  &  9.64  &$\pm$ 0.43  &$\pm$ 0.39 \\
70  & 7.04 &$\pm$ 0.55 &$\pm$ 0.32   & 6.64  &$\pm$ 0.49  &$\pm$ 0.27   & 6.59  &$\pm$ 0.35  &$\pm$ 0.22  &  9.07  &$\pm$ 0.49  &$\pm$ 0.41 \\
80  & 6.58 &$\pm$ 0.6 &$\pm$ 0.33    & 6.33  &$\pm$ 0.53  &$\pm$ 0.27   & 6.22  &$\pm$ 0.38  &$\pm$ 0.23  &  8.58  &$\pm$ 0.53  &$\pm$ 0.46 \\
90  & 6.72 &$\pm$ 0.77 &$\pm$ 0.36   & 6.83  &$\pm$ 0.74  &$\pm$ 0.27   & 6.53  &$\pm$ 0.51  &$\pm$ 0.23  &  9.02  &$\pm$ 0.71  &$\pm$ 0.51 \\
100 & 5.88 &$\pm$ 0.75 &$\pm$ 0.35   & 6.82  &$\pm$ 0.87  &$\pm$ 0.28   & 6.02  &$\pm$ 0.54  &$\pm$ 0.23  &  8.33  &$\pm$ 0.75  &$\pm$ 0.48 \\
110 & 5.90 &$\pm$ 0.87 &$\pm$ 0.44   & 6.76  &$\pm$ 1.05  &$\pm$ 0.28   & 6.01  &$\pm$ 0.64  &$\pm$ 0.28  &  8.25  &$\pm$ 0.88  &$\pm$ 0.53 \\
120 & 5.74 &$\pm$ 0.95 &$\pm$ 0.38   & 6.34  &$\pm$ 1.20  &$\pm$ 0.31   & 5.72  &$\pm$ 0.71  &$\pm$ 0.26  &  7.93  &$\pm$ 0.98  &$\pm$ 0.52 \\
130 & 5.76 &$\pm$ 1.12 &$\pm$ 0.45   & 7.22  &$\pm$ 1.72  &$\pm$ 0.30   & 5.95  &$\pm$ 0.89  &$\pm$ 0.31  &  8.31  &$\pm$ 1.25  &$\pm$ 0.55 \\
140 & 5.23 &$\pm$ 1.1 &$\pm$ 0.66    & 8.04  &$\pm$ 2.17  &$\pm$ 0.56   & 5.62  &$\pm$ 0.93  &$\pm$ 0.50  &  7.94  &$\pm$ 1.31  &$\pm$ 0.76 \\
150 & 5.58 &$\pm$ 1.4 &$\pm$ 0.50    & 7.40  &$\pm$ 2.37  &$\pm$ 0.76   & 5.70  &$\pm$ 1.13  &$\pm$ 0.40  &  8.15  &$\pm$ 1.62  &$\pm$ 0.67 \\
160 & 4.99 &$\pm$ 1.35 &$\pm$ 0.47   & 5.17  &$\pm$ 1.72  &$\pm$ 0.48   & 4.83  &$\pm$ 1.01  &$\pm$ 0.36  &  6.92  &$\pm$ 1.44  &$\pm$ 0.56 \\
170 & 6.19 &$\pm$ 2.02 &$\pm$ 0.70   & 5.30  &$\pm$ 2.09  &$\pm$ 0.59   & 5.53  &$\pm$ 1.39  &$\pm$ 0.53  &  7.97  &$\pm$ 2.00  &$\pm$ 0.84 \\
180 & 6.42 &$\pm$ 2.17 &$\pm$ 0.57   & 5.72  &$\pm$ 2.54  &$\pm$ 0.86   & 5.86  &$\pm$ 1.57  &$\pm$ 0.55  &  8.38  &$\pm$ 2.24  &$\pm$ 1.25 \\
190 & 6.9  &$\pm$ 2.5 &$\pm$ 0.94    & 5.70  &$\pm$ 2.65  &$\pm$ 0.84   & 6.04  &$\pm$ 1.72  &$\pm$ 0.72  &  8.43  &$\pm$ 2.40  &$\pm$ 1.46 \\
\hline \hline
\end{tabular}
\caption{Measured \rjets in the electron, muon and combined channels.  The
extrapolation to a common fiducial region for the combination decreases the
value of the ratio for both channels.  Values are reported with statistical
and systematic uncertainties, respectively.
 }
\label{tab:results}
\end{center}
\end{table*}

\section{Conclusions}
\label{sec:conclusions}
This Letter presents a first measurement of the ratio of the production
cross sections of the gauge bosons \Wboson and \Zboson in association with
exactly one jet.  Results are presented as a function of jet \pt threshold,
in both the electron and muon decay modes of the \Wboson and \Zboson vector
bosons. The measurement was corrected for all detector effects back to the
particle level and presented within a fiducial phase space for electrons and
muons. The ratio was measured to be $8.73 \pm 0.30 \,\mathrm{(stat)} \pm
0.40 \,\mathrm{(syst)}$ in the electron channel, and $8.49 \pm 0.23
\,\mathrm{(stat)} \pm 0.33 \,\mathrm{(syst)}$ in the muon channel at a jet
\pt threshold of $30 \GeV$ (Table \ref{tab:results}). Results have also been
extrapolated to $|\eta| <2.5$ and combined, yielding $8.29 \pm 0.18
\,\mathrm{(stat)} \pm 0.28 \,\mathrm{(syst)}$, and extrapolated to the full
phase of the boson and combined giving $10.13 \pm 0.22 \,\mathrm{(stat)} \pm
0.45 \,\mathrm{(syst)}$. The design of the measurement allows a cancellation
of many theoretical and systematic uncertainties.  These results are
provided as a function of jet \pt threshold from 30 to 200 GeV, exploring
the transition region of electroweak scale breaking in the perturbative jet
production.  This measurement builds the foundations of a high precision
test of the Standard Model, and provides model-independent sensitivity to
new physics coupling to leptons and jets.  Comparisons with LO and NLO
perturbative QCD predictions were made and found to be in agreement with
data over the jet \pT threshold range covered by this measurement.

\section{Acknowledgments}

We wish to thank CERN for the efficient commissioning and operation of the
LHC during this initial high-energy data-taking period as well as the
support staff from our institutions without whom ATLAS could not be operated
efficiently.

We acknowledge the support of ANPCyT, Argentina; YerPhI, Armenia; ARC,
Australia; BMWF, Austria; ANAS, Azerbaijan; SSTC, Belarus; CNPq and FAPESP,
Brazil; NSERC, NRC and CFI, Canada; CERN; CONICYT, Chile; CAS, MOST and
NSFC, China; COLCIENCIAS, Colombia; MSMT CR, MPO CR and VSC CR, Czech
Republic; DNRF, DNSRC and Lundbeck Foundation, Denmark; ARTEMIS, European
Union; IN2P3-CNRS, CEA-DSM / IRFU, France; GNAS, Georgia; BMBF, DFG, HGF,
MPG and AvH Foundation, Germany; GSRT, Greece; ISF, MINERVA, GIF, DIP and
Benoziyo Center, Israel; INFN, Italy; MEXT and JSPS, Japan; CNRST, Morocco;
FOM and NWO,Netherlands; RCN,Norway; MNiSW,Poland; GRICES and FCT, Portugal;
MERYS (MECTS), Romania; MES of Russia and ROSATOM, Russian Federation; JINR;
MSTD, Serbia; MSSR, Slovakia; ARRS and MVZT, Slovenia; DST / NRF, South
Africa; MICINN, Spain; SRC and Wallenberg Foundation, Sweden; SER, SNSF and
Cantons of Bern and Geneva, Switzerland; NSC, Taiwan; TAEK, Turkey; STFC,
the Royal Society and Leverhulme Trust, United Kingdom; DOE and NSF, United
States of America.

The crucial computing support from all WLCG partners is acknowledged
gratefully, in particular from CERN and the ATLAS Tier-1 facilities at
TRIUMF (Canada), NDGF (Denmark, Norway, Sweden), CC-IN2P3 (France),
KIT/GridKA (Germany), INFN-CNAF (Italy), NL-T1 (Netherlands), PIC (Spain),
ASGC (Taiwan), RAL (UK) and BNL (USA) and in the Tier-2 facilities
worldwide.

\bibliographystyle{model1-num-names}
\bibliography{rjets_plb}

\clearpage 
\onecolumn 

\begin{flushleft}
{\Large The ATLAS Collaboration}

\bigskip

G.~Aad$^{\rm 48}$,
B.~Abbott$^{\rm 111}$,
J.~Abdallah$^{\rm 11}$,
A.A.~Abdelalim$^{\rm 49}$,
A.~Abdesselam$^{\rm 118}$,
O.~Abdinov$^{\rm 10}$,
B.~Abi$^{\rm 112}$,
M.~Abolins$^{\rm 88}$,
H.~Abramowicz$^{\rm 153}$,
H.~Abreu$^{\rm 115}$
% ...
\bigskip

% ...
$^{10}$ Institute of Physics, Azerbaijan Academy of Sciences, Baku, Azerbaijan\\
$^{11}$ Institut de F\'isica d'Altes Energies and Departament de F\'isica de la Universitat Aut\`onoma  de Barcelona and ICREA, Barcelona, Spain\\
% ...
$^{48}$ Fakult\"{a}t f\"{u}r Mathematik und Physik, Albert-Ludwigs-Universit\"{a}t, Freiburg i.Br., Germany\\
$^{49}$ Section de Physique, Universit\'e de Gen\`eve, Geneva, Switzerland\\
% ...
$^{88}$ Department of Physics and Astronomy, Michigan State University, East Lansing MI, United States of America\\
% ...
$^{111}$ Homer L. Dodge Department of Physics and Astronomy, University of Oklahoma, Norman OK, United States of America\\
$^{112}$ Department of Physics, Oklahoma State University, Stillwater OK, United States of America\\
% ...
$^{115}$ LAL, Univ. Paris-Sud and CNRS/IN2P3, Orsay, France\\
% ...
$^{118}$ Department of Physics, Oxford University, Oxford, United Kingdom\\
% ...
$^{153}$ Raymond and Beverly Sackler School of Physics and Astronomy, Tel Aviv University, Tel Aviv, Israel\\
% ...
\end{flushleft}


\end{document}